\documentclass[twocolumn,showpacs,aps,prd,superscriptaddress,floatfix]{revtex4-1}
\usepackage{graphicx}  
\usepackage{dcolumn}   
\usepackage{bm}        
\usepackage{amssymb}   
\usepackage{amsfonts}  
\usepackage{hyperref}
\usepackage{amsmath}
\usepackage{mathtools}
\usepackage{xspace}           

\def\gev     {\ensuremath{\mathrm{GeV}}\xspace}
\def\gevc    {\ensuremath{\mathrm{GeV}/c}\xspace}
\def\gevcc   {\ensuremath{\mathrm{GeV}/c^2}\xspace}

\def\mevcc   {\ensuremath{\mathrm{MeV}/c^2}\xspace}

\def\simge{\mathrel{
   \rlap{\raise 0.511ex \hbox{$>$}}{\lower 0.511ex \hbox{$\sim$}}}}
\def\simle{\mathrel{
   \rlap{\raise 0.511ex \hbox{$<$}}{\lower 0.511ex \hbox{$\sim$}}}}

\begin{document}
\title{\bf Search for a light CP-odd Higgs boson in the radiative decays of {\boldmath $J/\psi$}}

\author{
  \begin{small}
    \begin{center}
      M.~Ablikim$^{1}$, M.~N.~Achasov$^{9,f}$, X.~C.~Ai$^{1}$,
      O.~Albayrak$^{5}$, M.~Albrecht$^{4}$, D.~J.~Ambrose$^{44}$,
      A.~Amoroso$^{49A,49C}$, F.~F.~An$^{1}$, Q.~An$^{46,a}$,
      J.~Z.~Bai$^{1}$
% ...
      \\
      \vspace{0.2cm}
      (BESIII Collaboration)\\
      \vspace{0.2cm} {\it
        $^{1}$ Institute of High Energy Physics, Beijing 100049, People's Republic of China\\
% ...
$^{4}$ Bochum Ruhr-University, D-44780 Bochum, Germany\\
        $^{5}$ Carnegie Mellon University, Pittsburgh, Pennsylvania 15213, USA\\
% ...
$^{9}$ G.I. Budker Institute of Nuclear Physics SB RAS (BINP), Novosibirsk 630090, Russia\\
% ...
$^{44}$ University of Rochester, Rochester, New York 14627, USA\\
% ...
$^{46}$ University of Science and Technology of China, Hefei 230026, People's Republic of China\\
% ...
$^{49}$ (A)University of Turin, I-10125, Turin, Italy; (B)University of Eastern Piedmont, I-15121, Alessandria, Italy; (C)INFN, I-10125, Turin, Italy\\
% ...
$^{a}$ Also at State Key Laboratory of Particle Detection and Electronics, Beijing 100049, Hefei 230026, People's Republic of China\\
% ...
$^{f}$ Also at the Novosibirsk State University, Novosibirsk, 630090, Russia\\
% ...
      }\end{center}
    \vspace{0.4cm}
  \end{small}
}

\begin{abstract}
  We search for a light Higgs boson $A^0$ in the fully reconstructed
  decay chain of $J/\psi \rightarrow \gamma A^0$, $A^0 \rightarrow
  \mu^+\mu^-$ using $(225.0\pm2.8)\times10^6$ $J/\psi$ events
  collected by the BESIII experiment. The $A^0$ is a hypothetical
  CP-odd light Higgs boson predicted by many extensions of the
  Standard Model including
two spin-0 doublets plus an extra singlet. We find no evidence for $A^0$ production and set $90\%$
  confidence-level upper limits on the product branching fraction
  $\mathcal{B}(J/\psi \rightarrow \gamma A^0) \times \mathcal{B}(A^0
  \rightarrow \mu^+\mu^-)$ in the range of $(2.8-495.3)\times
  10^{-8}$ for $0.212 \le m_{A^0} \le 3.0 \;\gevcc$. The new limits
  are  $5$ times below our previous results, and the nature of the $A^0$ is constrained to be mostly singlet.
\end{abstract}

\pacs{14.80.Ec, 14.60.St, 12.60.Fr, 14.60.Ef, 13.20.Fc, 14.65.Dw}

\maketitle

The radiative decays of the $J/\psi$ have long been identified as a way to search for new particles such as a light scalar, a pseudo-scalar Higgs boson \cite{Wilzek}, or a  light \hbox{spin-1} gauge boson \cite{Fayet81}.
In particular a light CP-odd pseudo-scalar may be present in various models
of physics beyond the Standard Model,
such as the
Next-to-Minimal Supersymmetric Standard Model (NMSSM)~\cite{NMSSM}. 
The NMSSM appends an additional singlet chiral
superfield to the Minimal Supersymmetric Standard Model (MSSM)~\cite{MSSM}, 
 in order to solve or alleviate the so-called \rm{\lq\lq little hierarchy problem\rq\rq}~\cite{littleHierarchy}. It has a rich Higgs sector containing 
three CP-even, two CP-odd and two charged Higgs bosons. The mass of the lightest CP-odd Higgs boson, $A^0$, may be less than twice the mass of the charmed quark.

The branching fraction of $V \rightarrow \gamma A^0$  ($V = \Upsilon, J/\psi$) is related to the Yukawa coupling of $A^0$ to the down or up  type of quark ($g_q^2$) through  \cite{Wilzek,Radiative_Quarkonium,P_Nason},  

\begin{equation}
\frac{\mathcal{B}(V \rightarrow \gamma A^0)}{\mathcal{B}(V \rightarrow l^+l^-)} = \frac{G_F m_q^2 g_q^2C_{QCD}}{\sqrt{2}\pi \alpha}\biggl(1-\frac{m_{A^0}^2}{m_V^2}\biggl)   
\label{Eq:fycoupling}
\end{equation}

\noindent where  $l \equiv e$ or $\mu$, $\alpha$ is the fine structure constant, $m_q$ the quark mass and $C_{QCD}$ the combined $m_{A^0}$ dependent QCD and relativistic corrections to $\mathcal{B}(V \rightarrow \gamma A^0)$~\cite{P_Nason} and the leptonic width of $\mathcal{B}(V \rightarrow l^+l^-)$~\cite{Barbieri}. The correction of first order in the strong coupling constant ($\alpha_{S}$) is as large as $30\%$~\cite{P_Nason} but comparable to the theoretical uncertainties~\cite{Beneke}. In the NMSSM, $g_c=\cos\theta_A/\tan \beta$ for the $c$-quark and $g_b=\cos \theta_A\tan\beta$ for the $b$-quark, where $\tan\beta$ is  the ratio of the expectation values of the up and down types of the Higgs doublets and $\cos\theta_A$ the fraction of the non-singlet component in the $A^0$ \cite{NMSSMconst,fayet};  $\cos \theta _A$ takes into account the doublet-singlet mixing and would be small for a mostly-singlet pseudoscalar \cite{Fayet81}.
The branching fraction of $J/\psi \rightarrow \gamma A^0$ could be in the
range of $\,10^{-9}\,$--$\,10^{-7}$~\cite{Gunion}, making it accessible at high intensity $e^+e^-$
collider experiments.

The BABAR
~\cite{Aubert:2009cp,Sanchez:2010bm,Lees:2011wb,vindy}, CLEO
~\cite{CLEO_Higgs0}, and CMS~\cite{CMS} experiments have performed
searches for $A^0$ in various decay processes and placed very strong
exclusion limits on $g_b$
~\cite{NMSSMconst,CMS,Lees:2011wb,vindy}. The BES\,III experiment, on the other hand, is sensitive to $g_c$. Existing constraints on $g_b$ give 
$\mathcal{B}(J/\psi\rightarrow A^0)  \times \mathcal{B}(A^0 \rightarrow \mu^+\mu^-) \,\simle  \,5 \times 10^{-7}\,\cot^4 \beta$, 
i.e.~$\simle 3 \times 10^{-8}$ for $\tan \beta \simge 2$ \cite{fayet}.
The search for the $A^0$ in $J/\psi$ experiments is particularly important at lower values of $\tan\beta$, typically for $\tan \beta \simle \,2$\,.

The BES\,III experiment has previously searched
for di-muon decays of light pseudoscalars, in the radiative decays of
$J/\psi$ using $\psi(2S)$ data, where the pion pair from $\psi(2S)
\rightarrow \pi^+\pi^- J/\psi$ was used to tag the $J/\psi$
events~\cite{BESIII_Higgs0}.  No candidates were found and exclusion
limits on $\mathcal{B}(J/\psi \rightarrow \gamma A^0) \times
\mathcal{B}(A^0 \rightarrow \mu^+\mu^-$) were set in the range of
$(0.4-21.0) \times 10^{-6}$ for  $0.212 \le m_{A^0} \le 3.0\;\gevcc$~\cite{BESIII_Higgs0}. 

This paper describes the search for a narrow $A^0$ signal in the fully
reconstructed process $J/\psi \rightarrow \gamma A^0$, $A^0 \rightarrow
\mu^+\mu^-$ using $(225.0\pm2.8)\times10^6$ $J/\psi$ events
collected by the BESIII experiment in 2009~\cite{bes3_cpc}. The same
amount of generic $J/\psi$ decays, generated by EvtGen~\cite{EVTGEV} where branching fractions of all the known decay processes are taken into account as mentioned in ~\cite{Bayesian},
is used for background studies. The $A^0$ is assumed to be a scalar or
pseudo-scalar particle with a very narrow decay width in comparison to
the experimental resolution~\cite{Fullana}.

BESIII is a general purpose spectrometer as described in
~\cite{bes3_nim}. It consists of four detector sub-components and has a
geometrical acceptance of $93\%$ of the total solid angle.  A helium
based ($40\%$ He, $60\%$ $C_3H_8$) 43 layer main drift chamber (MDC),
operating in a 1.0~T solenoidal magnetic field, is used to measure the
momentum of charged particles.  Charged particle identification (PID) is based
on the time-of-flight (TOF) measured by a scintillation based TOF system,
which has one barrel portion and two end-caps, and the energy loss
(d$E$/d$x$) in the tracking system. Photon and electron energies are
measured in a CsI(Tl) electromagnetic calorimeter (EMC), while muons
are identified using a muon counter (MUC) system containing nine
(eight) layers of resistive plate chamber counters interleaved with
steel in the barrel (end-cap) region.

We use simulated signal events with 23 different $A^0$ mass hypotheses ranging from 0.212 to 3.0 $\gevcc$ to study the detector acceptance and optimize
the event selection procedure. The decay of signal events
is simulated by the {\sc EvtGen} event generator~\cite{EVTGEV}, and a
phase-space  model is used for the $A^0 \rightarrow \mu^+\mu^-$ decay
and a $P$-Wave model for the decay $J/\psi \rightarrow \gamma A^0$. {\sc BABAYAGA 3.5}~\cite{babayaga} is
used to simulate the radiative Bhabha events, and {\sc PHOKHARA 7.0}~\cite{phokhara} to simulate initial state radiation
(ISR) processes of $e^+e^- \rightarrow \gamma \mu^+\mu^-$, $e^+e^-
\rightarrow \gamma \pi^+\pi^-$ and $e^+e^- \rightarrow \gamma
\pi^+\pi^-\pi^0$. A Monte Carlo (MC) simulation based  on the {\sc Geant4} package
~\cite{GEANT4} is used to determine the detector response and
reconstruction efficiencies.

We select events with exactly two oppositely charged tracks and at
least one good photon. The minimum energy of this photon is required
to be 25 MeV in the barrel region ($|\cos\theta| < 0.8$) and 50 MeV in
the end-cap region ($0.86 < |\cos\theta| < 0.92$). The EMC time is
also required to be in the range of $[0,14](\times 50)$ ns to suppress
electronic noise and energy deposits unrelated to the signal events.
Additional photons are allowed to be in the events. In order to reduce
the beam related backgrounds, charged tracks are required to have
their points of closest approach to the beam-line within $\pm 10.0$ $\mathrm{cm}$
from the interaction point in the beam direction and within $1.0 \;
\mathrm{cm}$ in the plane perpendicular to the beam. In order to
have a
reliable measurement in the MDC, they must be in the polar angle
region $|\cos\theta| < 0.93$. We suppress contamination by electrons
by requiring $E^{\mu}_\mathrm{cal}/p < 0.9$ $c$, where $E^{\mu}_\mathrm{cal}$ is the
energy deposited in the EMC by the showering particles and $p$ is the
incident momentum of the charged particles entering the
calorimeter. The angle between a photon and the nearest extrapolated
track in the EMC is required to be greater than 20 degrees (10
degrees) for $m_{A^0} \le 0.3\;\gevcc$ ($m_{A^0} > 0.3\;\gevcc$)
to remove bremsstrahlung photons.

We assign a muon mass hypothesis to the two charged tracks and require
that one of the charged tracks must be identified as a muon using the
muon PID system, which is based on the selection criteria: (1) $0.1 < E_\mathrm{cal}^{\mu} < 0.3\;\gev$, (2) the absolute value of the time
difference between TOF and expected muon time ($\Delta t^\mathrm{TOF}$) must
be less than 0.26 ns and (3) the penetration depth in MUC must be
greater than $(-40.0 +70\times p/(\gevc)$) $\mathrm{cm}$ for $0.5 \le p \le 1.1\;\gevc$
and 40~cm for $p > 1.1\;\gevc$. The two muon candidates are required
to meet at a common vertex to form the Higgs candidate. To improve the
mass resolution of the $A^0$ candidates,  a four-constraint (4C) kinematic
fit is performed with two charged tracks and each of the photons. If
there is more than one $\gamma \mu^+\mu^-$ candidate, the one with the minimum 4C $\chi^2$
is selected, and the $\chi^2$ is required to be less than $40$ to
suppress background contributions from $J/\psi \rightarrow \rho\pi$
and $e^+e^- \rightarrow \gamma \pi^+\pi^-\pi^0$. Fake photons are
eliminated by requiring the di-muon invariant mass, obtained from the
4C kinematic fit, to be less than $3.04\;\gevcc$. We further require
that one of the tracks must have the cosine of the muon helicity angle
($ \cos \theta_{\mu}^{hel}$), defined as the angle between the
direction of one of the muons and the direction of $J/\psi$ in the $A^0$ rest
frame, to be less than 0.92 to suppress the backgrounds peaking at $| \cos \theta_{\mu}^{hel}| \approx 1$.

\begin{figure}
\begin{center}
\includegraphics[width=0.5\textwidth]{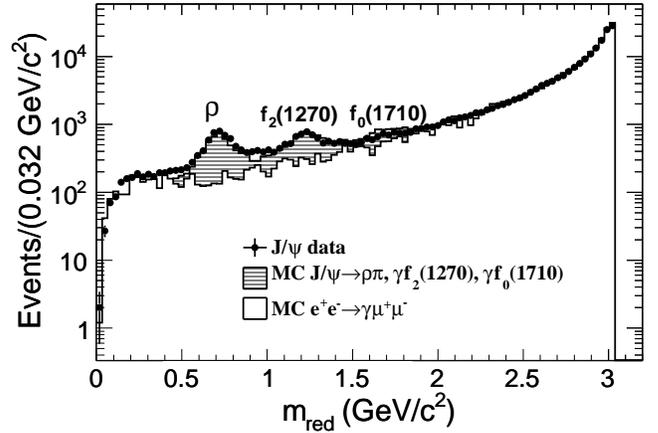}
\caption{Distribution of $m_{\rm red}$ for data (black points with
  error bars), together with the background predictions from the
  various MC samples, shown by a solid histogram and a histogram with
  horizontal pattern lines for the non-peaking and peaking
  backgrounds, respectively. The MC samples are normalized to the
  data. Three peaking components, corresponding to the $\rho$,
  $f_0(1270)$ and $f_0(1710)$ mesons, are observed in the data.}
\label{fig:mred}
\end{center}
\end{figure}

The above selection criteria select a total of 210,850 events in
$J/\psi$ data. Fig.~\ref{fig:mred} shows the distribution of the reduced
di-muon mass, $m_{\rm red} = \sqrt{m_{\mu^+\mu^-}^2 - 4 m_{\mu}^2}$,
of data together with the background predictions from various
simulated MC samples. $m_{\rm red}$ is equal to twice the muon
momentum in the $A^0$ rest frame, and is easier to model near
threshold than the di-muon invariant mass. The background is dominated
by the \rm{\lq\lq non-peaking\rq\rq} component of $e^+e^- \rightarrow
\gamma \mu^+\mu^-$ and the \rm{\lq\lq peaking\rq\rq} components of
$J/\psi \rightarrow \rho\pi$, $\gamma f_2(1270)$, and $\gamma
f_0(1710)$.

We perform a series of one dimensional unbinned extended maximum
likelihood (ML) fits to the $m_{\rm red}$ distribution to determine the
number of signal candidates as a function of $m_{A^0}$ in the interval
of $0.212 \le m_{A^0} \le 3.0\;\gevcc$. The likelihood function is a
combination of signal, continuum background and peaking background
contributions from $\rho$, $f_2(1270)$ and $f_0(1710)$ mesons. To handle the threshold-mass region and peaking backgrounds smoothly, the ML fit is done in intervals $0.002
\le m_{\rm red} \le 0.5\;\gevcc$ for $0.212 \le m_{A^0} \le 0.4\;
\gevcc$, $0.3 \le m_{\rm red} \le 0.65\;\gevcc$ for $0.4 <
m_{A^0} \le 0.6\;\gevcc$, $0.4 \le m_{\rm red} \le 1.1\;\gevcc$
for $0.6 < m_{A^0} \le 1.0\;\gevcc$, $0.9 \le m_{\rm red} \le 2.5\;
\gevcc$ for $1.0 < m_{A^0} \le 2.4\;\gevcc$ and $2.75 \le
m_{\rm red} \le 3.032\;\gevcc$ for $2.93 < m_{A^0} \le 3.0\;
\gevcc$. We use elsewhere the sliding intervals of $m-0.2 < m_{\rm
  red}< m+ 0.1\;\gevcc$, where $m$ is the mean of the $m_{\rm red}$
distribution.

We develop the probability density function (PDF) of signal and
backgrounds using the simulated MC events. The signal PDF in the
$m_{\rm red} $ distribution is parametrized by the sum of two
Crystal Ball (CB) functions~\cite{CB}. 
 The $m_{\rm red}$ resolution typically varies
from 2 to 12~\mevcc while the signal efficiency varies from $49\%$
to $33\%$ depending upon the momentum values of two muons at different
Higgs mass points. The signal efficiency and PDF parameters are
interpolated linearly between mass points. We use a polynomial
function $\sum_{l=1}^4 p_{l}m_{\rm red}^l$ to model the $m_{\rm red}$
distribution of non-peaking background in the threshold mass region of
$0.212 \le m_{A^0} \le 0.40\;\gevcc$, where $p_{l}$ are the
polynomial coefficients. This higher order polynomial function passes
through the origin when $m_{\rm red} =0$ and has enough degrees of
freedom to provide a threshold like behavior. We use a $2^\mathrm{nd}$ ($4^\mathrm{th}$ and $5^\mathrm{th}$) order Chebyshev
polynomial function to describe the $m_{\rm red}$ distribution
of non-peaking backgrounds  for $0.6 < m_{A^0} \le 1.0\;\gevcc$ and $2.40 < m_{A^0} < 2.75\;\gevcc$ ($2.85 \le m_{A^0} \le 2.93\;\gevcc$ and $2.93 < m_{A^0} \le 3.0\;\gevcc$, respectively) regions. For the remaining mass regions, we use a $3^\mathrm{rd}$ order Chebyshev polynomial function.

The $m_{\rm red}$ distribution of $\rho$ background is described by a
`Cruijff' function with a common peak position ($\mu$),
independent left and right widths ($\sigma_{LR}$), and non-Gaussian
tails ($\alpha_{L,R}$), whose parameters are determined from the MC
$J/\psi \rightarrow \rho\pi$ event sample. The `Cruijff'
function is defined as
\begin{equation}
 f_{L,R}(m_{\rm red})=  \exp[-(m_{\rm red}-\mu)^2/(2\sigma_{L,R}^2+\alpha_{L,R}(m_{\rm red}-\mu)^2)]. 
\label{Eq:cruijff}
\end{equation} 

\noindent The $f_2(1270)$ and $f_0(1710)$ peaking backgrounds are
described by the sum of two CB functions using parameters determined
from MC samples of $J/\psi \rightarrow \gamma X$, $X \rightarrow
\pi^+\pi^-$ decays, where $X=f_2(1270)$ and $f_0(1710)$ mesons.

\begin{figure}
\begin{center}
\includegraphics[width=0.5\textwidth]{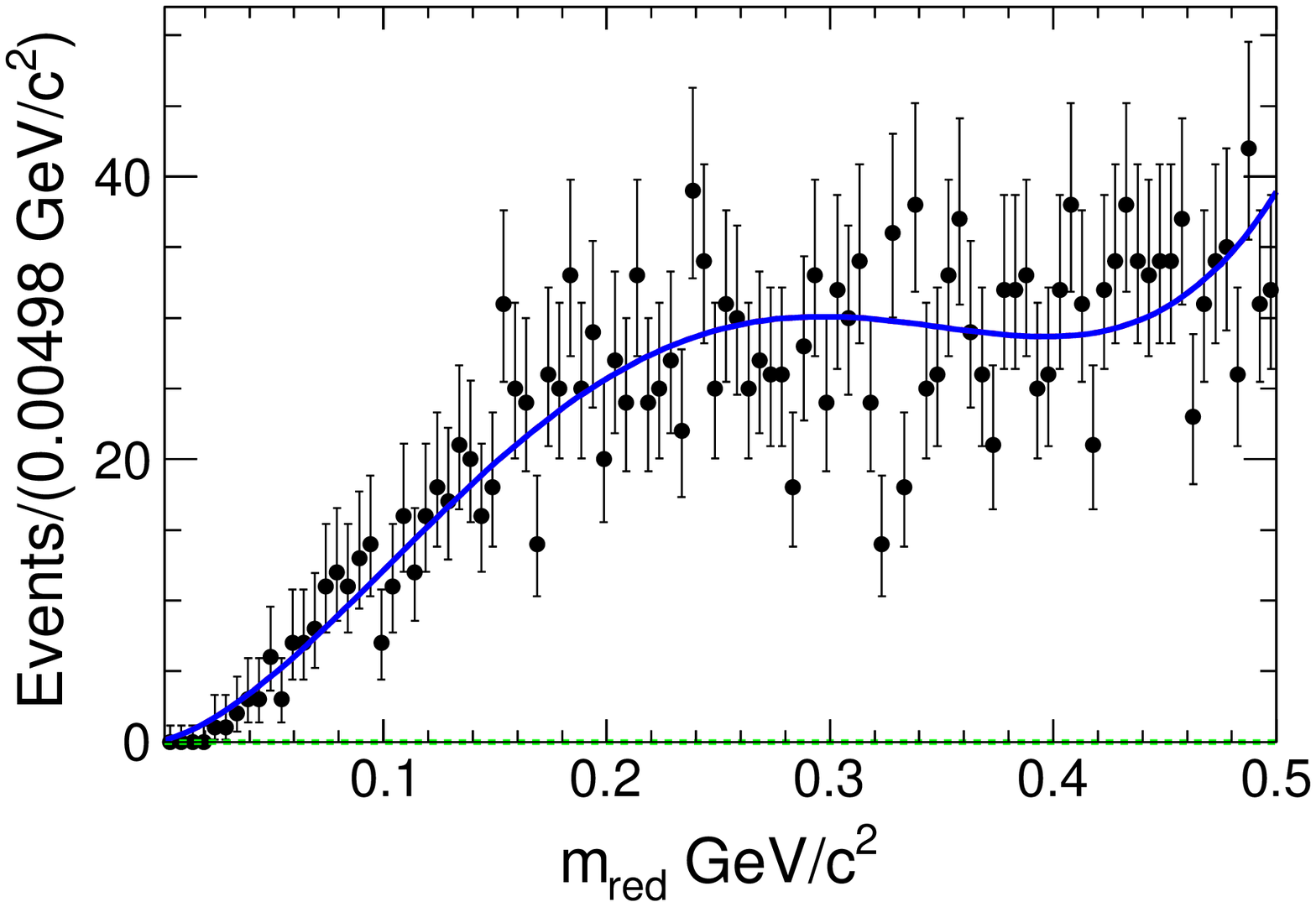}
\includegraphics[width=0.5\textwidth]{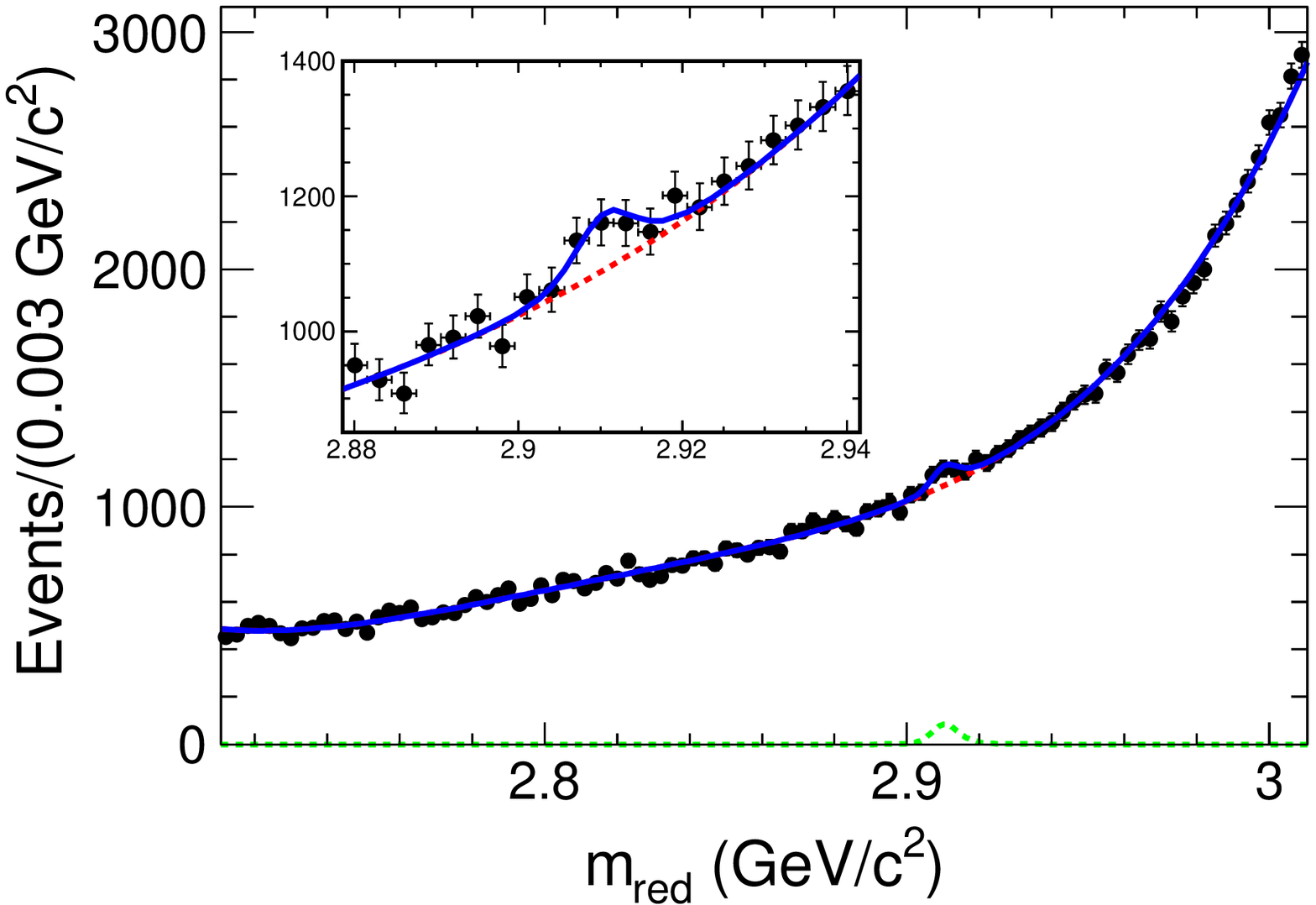}
\caption{(color online) Plot of the fit to the $m_{\rm red}$
  distribution for (top) $m_{A^0} = 0.212$ $\mathrm{GeV}/c^2$ and (bottom)  $m_{A^0} = 2.918$ $\mathrm{GeV}/c^2$. The contribution of non-peaking
  background is shown by a red dashed line, the signal PDF by a
  green dotted line (seen only in the bottom figure) and total PDF by a blue solid line. Due to limited statistics in the low-mass region as shown in the top figure, we allow the signal events to be floated for positive $N_{sig}$ only during the fit. The inlay in  the upper left of Fig.~(bottom) displays an enlargement of the $m_{\rm red}$ region
  between 2.88 and $2.94$ $\mathrm{GeV}/c^2$. The largest upward local significance is observed to be $3.42\sigma$ at $m_{A^0} = 2.918$ $\mathrm{GeV}/c^2$ point. }
\label{fig:proj}
\end{center}
\end{figure}

\begin{figure}
\begin{center}
\includegraphics[width=0.5\textwidth]{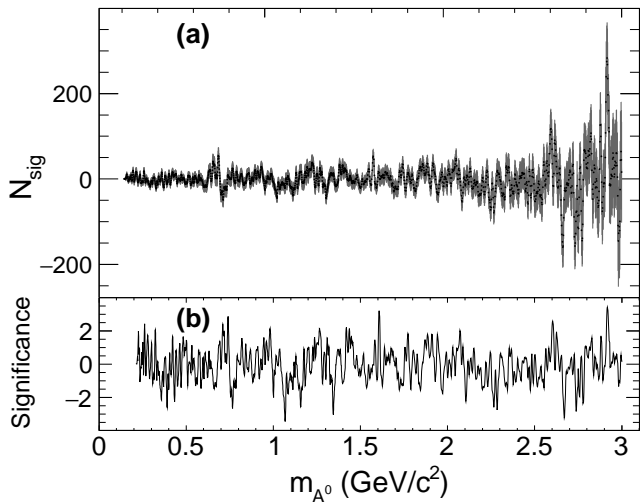}
\caption{(a) Number of signal events ($N_\mathrm{sig}$) and (b) signal
  significance ($\mathcal{S}$) obtained from the fit as a function of
  $m_{A^0}$.  }
\label{fig:yield}
\end{center}
\end{figure}

\begin{figure}
\begin{center}
\includegraphics[width=0.5\textwidth]{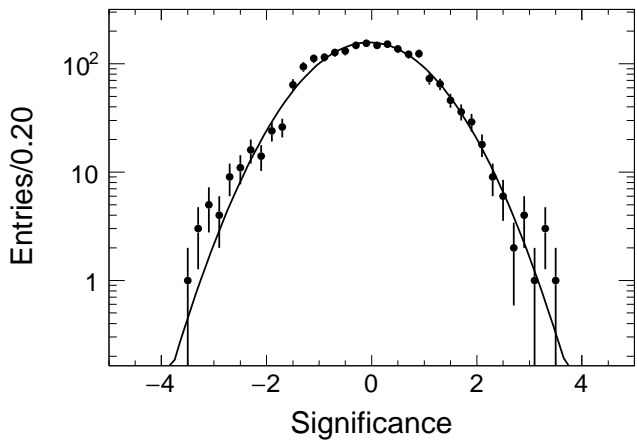}
\caption{Histogram of the statistical significance $\mathcal{S}$
  obtained from the fit at 2,035 $m_{A^0}$ points, together with the
  expected $\mathcal{S}$ distribution in the absence of signal, which
  is shown by the solid curve.}
\label{fig:signifdist}
\end{center}
\end{figure}

We search for a narrow resonance in steps of $1.0\;\mevcc$ in the mass
range of $0.22 \le m_{A^0} \le 1.50\;\gevcc$ and $2.0\;\mevcc$ for
other mass regions, resulting in a total of 2,035 $m_{A^0}$ points. The shapes
of the signal and the peaking background PDFs are fixed while the
non-peaking background PDF shape, and the numbers of signal, peaking
and non-peaking background events are left free in the fit. The plots of the fit to the $m_{\rm red}$ distribution for selected 
$m_{A^0}$ points are shown in
Fig.~\ref{fig:proj}. Fig.~\ref{fig:yield} shows signal event ($N_\mathrm{sig}$) and the
statistical significance, defined as $\mathcal{S} = {\rm
  sign}(N_\mathrm{sig})\sqrt{-2\ln(\mathcal{L}_0/\mathcal{L}_\mathrm{max})}$, as a
function of $m_{A^0}$, where $\mathcal{L}_\mathrm{max}$ ($\mathcal{L}_0$) is
the maximum likelihood value for a fit with number of signal events being
floated (fixed at zero). The distribution of $\mathcal{S}$ is expected
to follow the normal distribution under the null hypothesis,
consistent with the distribution in Fig.~\ref{fig:signifdist}. The
largest upward local significance is $3.42\sigma$ at $m_{A^0} = 2.918 \;
\gevcc$.

We repeat the
search  using a polynomial
function $\sum_{l=1}^5 p_{l}m_{\rm red}^l$ for $m_{A^0} \le 0.4\;\gevcc$ and an alternative higher order Chebyshev
polynomial function for other mass regions to model the non-peaking background. The difference between the absolute values of two $N_ \mathrm{sig} $  is considered as an additive systematic uncertainty at each mass point. An additive uncertainty reduces the significance of any
observed signal and does not scale with the number of reconstructed
signal events.

We study a large ensemble of pseudo-experiments, based on the aforementioned PDFs, to validate
the fit procedure and compute the bias of the ML fit.  The bias
arises due to the imperfections in modeling the signal PDFs and the
low statistics of the ML estimate.  The value of the fit bias is found to be 0.21
events and considered to be an additive systematic
uncertainty.  We further use the pseudo-experiments to estimate the
probability of observing a fluctuation of $\mathcal{S} \ge
3.42\sigma$, which is found to be $26.0\%$. The corresponding
global significance of such an excess anywhere in the full $m_{A^0}$ range is $0.64\sigma$; we therefore conclude that
no evidence of $A^0$ production is found at any mass points.

The uncertainty due to fixed signal and tail PDF parameters used for
the $\rho$, $f_2(1270)$ and $f_0(1710)$ peaking backgrounds in data,
is observed to be $(0.0-1.64)$ events after varying each parameter
within its statistical uncertainties while taking correlations between
the parameters into account. The mean and sigma values of the peaking
backgrounds are corrected using a high statistics control sample of
the same decay process in which all the selection criteria, developed
in this work, are applied except that of the penetration depth in
MUC. We assign $50\%$ of the relative difference in resolution values
of peaking backgrounds between data and MC as a systematic
uncertainty, which is considered as a source of multiplicative
systematic uncertainty. Multiplicative uncertainties scale with the
number of reconstructed signal events and do not reduce the
significance of any observed signal, but degrade the upper limit
values. They arise due to the reconstruction efficiency, the
uncertainty in the number of $J/\psi$ mesons ($1.3\%$),  muon tracking efficiency ($1.0\%$ per track) and
resolution of peaking backgrounds ($1.2\%$ for the $\rho$ resonance
and $6.52\%$ for $f_2(1270)$ and $f_0(1710)$ resonances).

We measure the photon reconstruction systematic uncertainty to be
better than $1.0\%$ using a $e^+e^- \rightarrow \gamma \mu^+\mu^-$
sample in which the ISR photon momentum is estimated using the
four-momenta of two charged tracks~\cite{photon_efficiency}. We use a
$J/\psi \rightarrow \mu^+\mu^-(\gamma)$ control sample, where one
track is tagged with tight muon PID and photons are produced via final state radiation, to study the systematic
uncertainty associated with the muon PID ($(4.0-5.73)\%$),
$\chi_{4C}^2$ ($1.56\%$) and the $\cos \theta_{\mu}^{hel}$ ($0.34\%$)
requirements. The final muon PID uncertainty also takes into account
the fraction of events with one track or two tracks identified as
muons, which is obtained from the signal MC. The total multiplicative
systematic uncertainty varies in the range of $(5.03 - 9.20)\%$
depending on $m_{A^0}$.
     
\begin{figure}
\begin{center}
\includegraphics[width=0.5\textwidth]{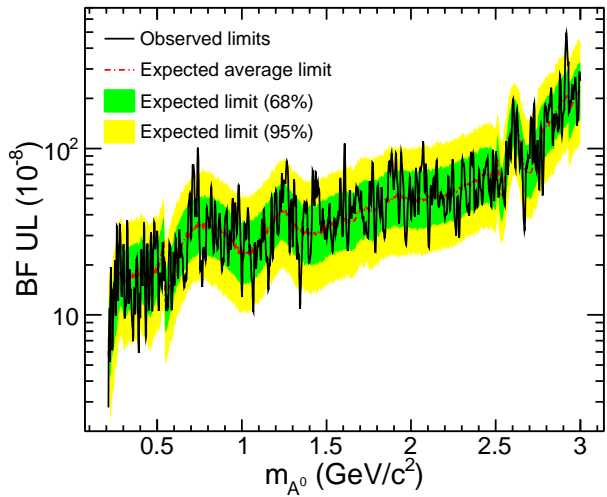}
\caption{(color online) The $90\%$ C.L. upper limits (UL) on the
  product branching fractions $\mathcal{B}(J/\psi \rightarrow \gamma
  A^0) \times \mathcal{B}(A^0 \rightarrow \mu^+\mu^-)$ as a function
  of $m_{A^0}$ including all the uncertainties (solid line), together with expected
  limits computed using a large number of pseudo-experiments. The
  inner and outer bands include statistical uncertainties only and
  contain $68\%$ and $95\%$ of the expected limit values. The average  dashed
  line in the center of the inner band is the expected average upper
  limit of 1600 pseudo-experiments. A better sensitivity in the mass region of $0.212 \le m_{A^0} \le 0.22\;\gevcc$ is achieved due to almost negligible backgrounds as seen in Fig.~\ref{fig:proj}~(top).  }
\label{fig:limit}
\end{center}
\end{figure}

We compute the $90\%$ confidence-level (C.L.) upper limits on the
product branching fractions of $\mathcal{B}(J/\psi \rightarrow \gamma
A^0) \times \mathcal{B}(A^0 \rightarrow \mu^+\mu^-)$ as a function of
$m_{A^0}$ using a Bayesian method ~\cite{Bayesian}.  The systematic uncertainty is incorporated by convolving the negative log likelihood (NLL)
versus branching fraction curve with a Gaussian distribution having
a width equal to the systematic uncertainty. The limits range between
$(2.8-495.3)\times 10^{-8}$ for the Higgs mass region of $0.212 \le
m_{A^0} \le 3.0\;\gevcc$ depending on the $A^0$ mass points, as shown in 
Fig.~\ref{fig:limit}. 

\begin{figure}
\begin{center}
\includegraphics[width=0.5\textwidth]{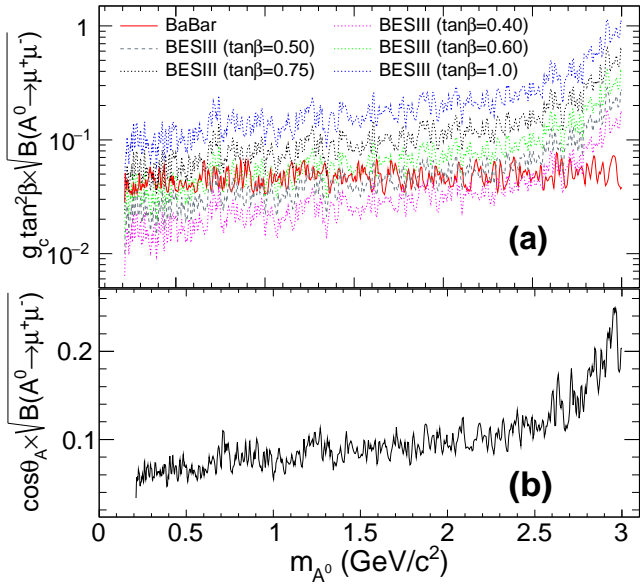}
\caption{(color online) (a) The $90\%$ C.L. upper limits on $g_b(=g_c  \tan^2\beta)\times \sqrt{\mathcal{B}(A^0 \rightarrow \mu^+\mu^-)}$ for the BABAR~\cite{vindy} and BESIII measurements and (b)  $\,\cos \theta_A (=|\sqrt{g_bg_c}|)\times   \, \sqrt{\mathcal{B}(A^0 \rightarrow \mu^+\mu^-)}$\ as a function of $m_{A^0}$. We compute $g_c  \tan^2\beta\times \sqrt{\mathcal{B}(A^0 \rightarrow \mu^+\mu^-)}$ for different values of $\tan\beta$ to compare our results with the BABAR measurement~\cite{vindy}. }
\label{fig:yukawa}
\end{center}
\end{figure}

 We also compute $g_b(=g_c\tan^2\beta)\times \sqrt{\mathcal{B}(A^0 \rightarrow \mu^+\mu^-)}$~\cite{fayet} for different values of $\tan\beta$ using Equation~\ref{Eq:fycoupling} to compare our results with the BABAR measurement~\cite{vindy}. This new result seems to be better than the BABAR measurement~\cite{vindy} in the low-mass region for $\tan\beta \le 0.6$ (Fig.~\ref{fig:yukawa} (a)). Our results are thus complementary to those obtained
by considering the $b$-quark~\cite{NMSSMconst,vindy}. Both types of constraints
may then be combined so as to provide,  {\it  independently of $\,\tan\beta$},  an upper limit on 
$\,\cos \theta_A (=|\sqrt{g_b g_c}|) \times   \, \sqrt{\mathcal{B}(A^0 \rightarrow \mu^+\mu^-)}$\ computed using the method of Ref.~\cite{fayet}, as a function of $m_{A^0}$, as shown in Fig.~\ref{fig:yukawa} (b)\,. This combined limit  varies in the range of $0.034-0.249$ for $0.212 \le m_{A^0} \le 3.0$ $\gevcc$.

In summary, we find no significant signal for a light Higgs boson in
the radiative decays of $J/\psi$ and set $90\%$ C.L. upper limits on
the product branching fraction of $\mathcal{B}(J/\psi \rightarrow
\gamma A^0) \times \mathcal{B}(A^0 \rightarrow \mu^+\mu^-)$ in the
range of $(2.8-495.3)\times 10^{-8}$ for $0.212 \le m_{A^0} \le 3.0\;
\gevcc$. This result, a factor of $5$ times improvement  over the previous BESIII measurement~\cite{BESIII_Higgs0}, is in agreement with the theoretical expectation \,$\simle  \,5 \times 10^{-7}\,\cot^4 \beta$ from \cite{fayet}, but better than the  BABAR measurement~\cite{vindy}  in the low-mass region for the $\tan\beta \le 0.6$. The combined limits on  $\,\cos \theta_A \times   \, \sqrt{\mathcal{B}(A^0 \rightarrow \mu^+\mu^-)}$\ for the BABAR~\cite{vindy} and BESIII measurements reveal that the $A^0$ is constrained to be mostly singlet.

\section{Acknowledgement}
The authors wish to thank Pierre Fayet for helpful discussions of new physics models. The BESIII collaboration thanks the staff of BEPCII and the IHEP
computing center for their strong support. This work is supported in
part by National Key Basic Research Program of China under Contract
No. 2015CB856700; National Natural Science Foundation of China (NSFC)
under Contracts Nos. 11125525, 11235011, 11322544, 11335008, 11425524;
the Chinese Academy of Sciences (CAS) Large-Scale Scientific Facility
Program; the CAS Center for Excellence in Particle Physics (CCEPP);
the Collaborative Innovation Center for Particles and Interactions
(CICPI); Joint Large-Scale Scientific Facility Funds of the NSFC and
CAS under Contracts Nos. 11179007, U1232201, U1332201; CAS under
Contracts Nos. KJCX2-YW-N29, KJCX2-YW-N45; 100 Talents Program of CAS;
National 1000 Talents Program of China; INPAC and Shanghai Key
Laboratory for Particle Physics and Cosmology; German Research
Foundation DFG under Contract No. Collaborative Research Center
CRC-1044; Istituto Nazionale di Fisica Nucleare, Italy; Joint Funds of
the National Science Foundation of China under Contract No. U1232107;
Ministry of Development of Turkey under Contract No. DPT2006K-120470;
Russian Foundation for Basic Research under Contract No. 14-07-91152;
The Swedish Resarch Council; U. S. Department of Energy under
Contracts Nos. DE-FG02-04ER41291, DE-FG02-05ER41374, DE-SC0012069,
DESC0010118; U.S. National Science Foundation; University of Groningen
(RuG) and the Helmholtzzentrum fuer Schwerionenforschung GmbH (GSI),
Darmstadt; WCU Program of National Research Foundation of Korea under
Contract No. R32-2008-000-10155-0.

\end{document}